\documentclass[fleqn,10pt]{wlscirep}
\usepackage{soul}
\usepackage{xcolor}
\newcommand{\KK}[1]{\textcolor{black}{#1}}
\newcommand{\wymiarsmall}{0.5\textwidth} 
\newcommand{\wymiarmediumone}{0.6\textwidth} 
\newcommand{\wymiarmediumtwo}{0.75\textwidth} 
\newcommand{\wymiarlarge}{0.95\textwidth} 
\newcommand{\wymiarlargetwo}{0.985\textwidth} 
\begin{document}
\title{Quench dynamics of Fano-like resonances\\ in the presence of the on-dot superconducting pairing}
\author[1,*]{Jan Bara\'nski}
\author[1]{Magdalena Bara\'nska}
\author[1]{Tomasz Zienkiewicz}
\author[2,3,*]{Konrad Jerzy Kapcia}
%
%
%
\affil[1]{Department of General Education, Polish Air Force University, ul. Dywizjonu 303 nr 35, 08521 Deblin, Poland} 
\affil[2]{Institute of Spintronics and Quantum Information, Faculty of Physics, Adam Mickiewicz University in Pozna\'n, ul. Uniwersytetu Pozna\'{n}skiego 2, 61614 Pozna\'{n}, Poland}
\affil[3]{Center for Free-Electron Laser Science CFEL, Deutsches Elektronen-Synchrotron DESY, Notkestr. 85, 22607 Hamburg, Germany}
\affil[*]{Correspondence should be addressed to j.baranski@law.mil.pl (J.B.) or konrad.kapcia@amu.edu.pl (K.J.K.).}
%
%
%
\begin{abstract}
We explore the electron dynamics of a system composed of double quantum dot embedded between metallic and superconducting leads in a ``T-shape'' geometry. 
In nanoscopic systems, where electron transfer between electrodes can be realized via different paths, interference effects play an important role. 
For double quantum dot system in the chosen geometry, interference of electrons transferred between electrodes via the interfacial quantum dot and electrons scattered on the side dot gives rise to Fano-like interference. 
If such a system is additionally coupled to a superconducting electrode, together with the well-understood Fano resonance an additional resonance appears on the opposite side of the Fermi level. 
In the recent work  [\href{https://doi.org/10.1038/s41598-020-59498-y}{ Sci. Rep. \textbf{10}, 2881 (2020)}], we showed that this resonance occurs solely as a result of the local pairing of non-scattered electrons with scattered ones. 
In this work, considering the quench dynamics, we explore how much time is required for formation of each of these resonances. 
In particular, (i) we analyze the charge oscillations between subsystems; (ii) we estimate the time required for each resonance to achieve stable equilibrium upon an abrupt change of interdot connection; (iii) we discuss a typical energy and time scales for experiments on similar architectures.
\end{abstract}
\flushbottom
\maketitle
\section*{Introduction}

Understanding of electron dynamics in nanoscopic systems is important issue for development of modern information processing devices.
To determine the upper limit of how fast such devices can perform logical operations, the system reaction time to a sudden change of parameters should be taken into account. 
Process in which one parameter of quantum system is changed much faster than time required for its thermalization is called a quantum quench \cite{Das2020}.

Equilibrium conditions analysis of charge transport through mesoscopic heterostructures turns out to be invaluable in the context of understanding of quantum interference effects and their coexistence with other many-body effects. In particular, the Fano resonance originating from interference of electrons transmitted simultaneously through broad and narrow energy levels has been extensively explored both on theoretical as well as experimental ground.
Interplay of the Fano-like interference with  the Coulomb blockade \cite{Lu_2006,Johnson}, the Kondo effect \cite{Bulka,Zitko,Sasaki}, topological states \cite{Schuray,Calle2020,Baraski2016,Xu_2022}, proximity-induced pairing \cite{Calle_2013,Calle_2017,Ramos_Andrade_2019,baranski.zienkiewicz.20,Bara_ski_2012,Bara_ski_2011} or electron-boson interaction  \cite{Baraski2015,Ma_2004,Gao_2008} are only a few examples. 
It was also established that sensitivity of Fano resonance for electron phase shift could be useful for probing of decoherences \cite{Baranski_2012_PRB,Rotter_2005,Barthaler}. 
Asymmetric Fano-line shapes have been also experimentally observed in various mesoscopic structures including double quantum dots \cite{Sasaki}, Aharonov-Bohm rings \cite{Kobayashi2003,Kobayashi} or quantum wires with a side-coupled quantum dot \cite{Sato,Kobayashi_2004}.

One particularly interesting experiment has been conducted by K. Kobayashi et al. \cite{Kobayashi}. 
The authors realized controllable device in which the Fano effect can be electrostatically turned on and off. 
The architecture used by the authors is based on a ,,bridge'' concept, where quantum dot is placed between the electrodes and an additional ``bridge'' electrode couples source and drain directly. Conceptually such realization is the same as double slit experiment for electrons with exception that one ,,slit'' (namely QD) possess discrete energy levels, while bridge electrode has continuum of states. 
By electrostatic pinching on and off the bridge arm, the authors were able to tune the system from ballistic to interferometric regime. 
Although results presented in \cite{Kobayashi} are gathered in equilibrium conditions, the experiment seems to be a promising ansatz for analysis of time required for formation of the Fano-type interference.
Implementation of similar experiment using modern techniques which provide an insight into electron time domain with picosecond resolution \cite{Cavalieri2007,Hentschel2001,Kienberger2004} would allow for an experimental inspection of the Fano resonance formation time. 
This perspective encourages to theoretical approaches concerning dynamics of interference effects in quantum dots.

Transient effects in quantum dot hybrids in various configurations have already been studied theoretically by a few authors \cite{Pan2009,Kwapinski_2014,Taranko_2015,Moldoveanu}. 
Dynamics of superconducting (SS) proximity effect has been inspected as well in Refs. \cite{RT2021,Weymann1,Weymann2,Levy1,Levy2,Levy3} including  also topological phenomena \cite{Levy4}. 
Recently, our group has also contributed to such studies \cite{baranski.baranska.21}.
Dynamics of Fano-type interference, however, has been presented only in \KK{Refs.~\cite{Michaek2022,Baranski2023}}.
The aforementioned work\KK{s focus} on the case of two quantum dots coupled to non-superconducting/metallic electrodes in T-shape geometry. 
Steady state analysis of our group on Fano resonances in presence of superconductors shows that interplay of quantum interference with local pairing leads to unique effects such as anomalous Fano resonance \cite{baranski.zienkiewicz.20}. 
In this work we will analyze such effects upon quench dynamics. 
To our knowledge time resolved analysis of interference effects in presence of local pairing has not been conducted yet.

One of convenient systems for theoretical analysis of such interferences in nanoscopic systems is composed of two quantum dots deposited in such a way that one quantum dot ($QD_1$) is strongly coupled to electrodes, while the other one ($QD_2$) is only side attached to the first quantum dot. This is \KK{so-called} "T-shape" configuration. Realization of similar architecture have been done e.g. in \cite{Kobayashi_2004}. 
In this architecture lifetime of electron on $QD_1$ is relatively short and in accordance to Fermi golden rule its energy levels become broad. In contrary, due to lack of direct coupling of $QD_2$ to continuum of states, its energy levels remain quasidiscrete. 
Interference of electrons transmitted through broad level and those resonantly scattered on discrete level gives rise to well pronounced Fano-like resonances.
Asymmetric Fano-like lineshapes can be observed both in density of states of interfacial dot ($QD_1$) near energy of discrete level, as well as in differential conductivity for gate voltages close to narrow level energy.

\begin{figure}
        \centering
	\includegraphics[width=\wymiarsmall]{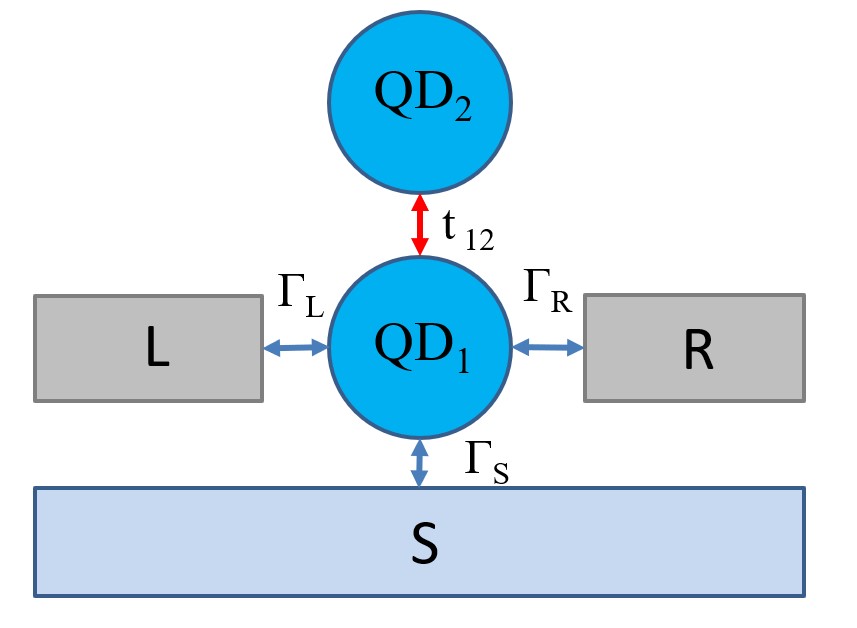}
	\caption{\textbf{The schematic illustration of the analyzed system.} 
        It consists of the double quantum dot (i.e., two quantum dots: $QD_1$ and $QD_2$) 
        embedded between two metallic ($L$, $R$) electrodes and one superconducting electrode ($S$).}
	\label{fig1a}
\end{figure}

In the present work, we inspect the dynamics of the interference effects when such system is additionally connected to superconducting electrode.
Thus, the full system considered here is composed of a double quantum dot deposited between two metallic ($L$ and $R$) and one superconducting ($S$) electrodes as schematically depicted in Figure \ref{fig1a}. 
Our goal is to estimate how much time is required for formation of Fano-like resonances upon establishing abrupt connection of interdot coupling $t_{12}$.

In our recent work on a similar architecture system, we showed that, if the local pairing is present, along the well-understood Fano resonance another feature on the opposite side of the Fermi level is formed~\cite{baranski.zienkiewicz.20}. 
The shape of this feature and the nature of its origin reach beyond traditional descriptions of the Fano-like resonances. 
It was shown that this anomalous Fano (AF) feature arises as a result of pairing of non-scattered electrons with scattered ones.
In particular, we considered a ``toy model'', where we assumed spin-polarized interdot tunneling. 
Such an assumption ensures that only one spin (say $\uparrow$) electrons are directly scattered on the side level and eventual resonant features in the other spin component ($\downarrow$-spin) originate solely from pairing with scattered ones. 
For such a model, in the spectrum of directly scattered electrons ($\uparrow$-spin), we observe only the ordinary Fano resonance while for electrons that are not directly scattered ($\downarrow$-spin) only the anomalous resonance appears. 
This shows that information about scattering of a given electron on the side level is transferred to another electron via the local pairing and AF resonance arises as a response \KK{to} scattering of opposite spin electron.
In a case of non-polarized tunneling, both electrons experience simultaneously direct scattering (leading to the emergence of the ordinary Fano (OF) resonance near $\epsilon_2$) and coupling to scattered (opposite spin) electrons leading to appearance of AF resonance near $-\epsilon_2$. 
Consequently, in the non-polarized case, both resonances are present in density of states and differential conductivity.

In the context of quench dynamics, an important question arises: \emph{Is there a difference in the timing of the formation of these two resonances?} 
On one hand, if one foresees the appearance of the AF features in the spectrum of a given spin electrons, the electron of opposite spin should be scattered on the side level firstly, and only then information about this event should be passed to the first electron giving rise to the AF resonance.
From this perspective, one could expect that the feature of direct scattering will be formed faster than its superconducting response. 
On the other hand, due to the local pairing, two electrons of opposite spin form one quantum object. 
If the superconducting-like correlations on the QD's are instantaneous, than any process that involve one electron should have its instantaneous effect on the other. 
Therefore, if information transfer between paired electrons is infinitely fast, there should not be any delay between formation of the OF and the AF resonances. 
In this regard, our analysis can be understood as answering the question \emph{whether the information transfer through local pairing have instantaneous character?}

In the present work, we explore the electron dynamics of a system composed of double quantum dot embedded between metallic and superconducting leads in a ``T-shape'' geometry.
Considering the quench dynamics, we explore how much time is required for formation of both the ordinary Fano resonance (associated with direct scattering of electrons) and the anomalous Fano resonant features (associated with indirect scattering of electrons due to the superconducting pairing). 
In particular, we analyze the charge oscillations between subsystems in the absence of normal electrodes.
The main part of the paper is devoted to estimation of the time required for both resonances to achieve stable equilibrium upon an abrupt change of interdot connection. 
Moreover, we discuss typical energy and time scales for experiments on similar architectures.

\section*{Model}

A setup investigated in this work and schematically depicted in Figure \ref{fig1a} can be described by the following Hamiltonian:
\begin{equation}
\label{eq:1.hamiltonian}
\hat{H}= \sum_{\beta=L,R,S} (\hat{H_{\beta}} + \hat{H}_{T \beta}) + \sum_{i=1,2} \hat{H}_{QDi} + \hat{H}_{t},
\end{equation}
where $\hat{H}_{\beta} $ correspond to  three external electrodes (two metallic $\beta=L,R$ and one superconducting $\beta=S$). 
Two single-level quantum dots and the tunnel coupling between them are represented by $\hat{H}_{QDi}= 
\sum_{ \sigma} \epsilon_{i} \hat{d}^{\dagger}_{i \sigma} \hat{d}_{i \sigma} + 
\sum_{ \sigma} U_{i} \hat{n}_{i \sigma} \hat{n}_{i \bar{\sigma}}$ and 
$\hat{H}_{t}=\sum_{\sigma} t_{12 \sigma} \left( \hat{d}^{\dagger}_{1 \sigma} \hat{d}_{2 \sigma} + 
\hat{d}^{\dagger}_{2 \sigma} \hat{d}_{1 \sigma}\right)$,  respectively ($\sigma = \uparrow, \downarrow$). 
$\hat{H}_{T \beta}$ stands for hybridization of each electrode to interfacial quantum dot ($QD_1$). 
We treat metallic reservoirs as free electron gas, thus for $\beta=L,R$ we have
$\hat{H}_{\beta} = \sum_{k,{\beta}, \sigma} \xi_{k {\beta} } \hat{c}^{\dagger}_{k {\beta} \sigma } \hat{c}_{k {\beta} \sigma }$, where electron energies $\xi_{k {\beta} }$ are measured with respect to chemical potentials $\mu_\beta$. 
Assuming isotropic energy gap ($\Delta_S$) of superconducting electrode, $\hat{H}_{\beta}$ for $\beta=S$ reads as 
$\hat{H}_{S}=\sum_{k,S, \sigma} \xi_{k S} \hat{c}^{\dagger}_{k S \sigma } \hat{c}_{k S \sigma} - \sum_k ( \Delta_S \hat{c}^{\dagger} _{k S \uparrow } \hat{c}^{\dagger}_{-k S \downarrow }+h.c.)$. 
The hybridization part is described by $\hat{H}_{T \beta}= 
\sum_{k, \sigma} V_{k \beta } \left(\hat{c}_{k \beta \sigma} \hat{d}^{\dagger}_{1 \sigma} + h.c.\right)$  
and represents the coupling of $QD_1$ to $\beta$ electrode.

Restricting our analysis to the energies deep inside the superconductor energy gap (i.e., $\Delta_S \rightarrow \infty$), we can neglect the existence of single-electron energy levels in the SC electrode. 
This, so-called superconducting atomic limit approximation, allows to reduce the influence of SC electrode on $QD_1$ by a static terms representing induced local pairing ($\hat{d}_{1 \uparrow} \hat{d}_{1 \downarrow} + h.c.$) \cite{Baraski2013,Tanaka2007,Oguri2004}. 
Studies on similar architecture in static conditions \cite{Baranski_2012_PRB,Bara_ski_2011,baranski.zienkiewicz.20} show that correlations have a marignal effect on appearance of interference features apart from two specific cases. 
First one, when energy level of a side dot is close to 0 and interference patterns coincide with Kondo resonance leading to the emergence of so-called Fano-Kondo resonance \cite{Zitko}. 
The second case is when Coulomb satellite state of $QD_2$ (with energy $\epsilon_2+U_{2}$) leads to appearance of additional interference structures \cite{Baranski_2012_PRB}. 
Transient effects in the Kondo regime were also studied in single-quantum dot systems, e.g., Refs. \cite{Cheng2015,Nghiem2017}.
We note that in experimental realizations correlations are usually much stronger than effective pairing and thus satellite state lies far beyond the considered energy scale. 
Let us underline that transient effects of laterally coupled quantum dots between normal and superconducting electrodes in both uncorrelated and correlated regimes was studied in Ref. \cite{Weymann2}. 
As it was pointed there, for the hybridization of quantum dots to  the SC electrode is stronger than correlations on the dots, the system is dominated by the Andreev scattering. 
In the mentioned work, even for correlations strength of  $U = 0.5 \Gamma_{S}$ ($U_1=U_2=U$, the Hartree-Fock-Bogoliubov decoupling scheme used), its influence on conductance was rather marginal and it was only limited to the influence on the final shape of transmittance and not the time of its evolution. 
In the case of strongly correlated system (i.e., $U=1.5 \Gamma_{S}$), two Andreev states form a single Lorentzian, however, the evolution in the time domain of all features remains the same.
Thus, one expects that the Coulomb repulsion, although it rearranges the static background transmittance, has rather a negligible effect on evolution of the resonant features in time.
Detailed analysis of this issue is out of the scope of the present work.

In this work, we focus solely on building up of interference effects in case when ordinary and anomalous Fano resonances are well separated (this is $\epsilon_2 \neq 0$). Therefore, in our calculations, we will omit onsite Coulomb interactions.
Consequently, constituents of the Hamiltonian representing the interfacial quantum dot, the SC electrode, and their mutual interactions can be rewritten in the following form:
\begin{eqnarray}
\hat{H}_{QD1}\!+\!\hat{H}_{S}\!+\!\hat{H}_{T S}\!=\!\sum_{\sigma}\! \epsilon_{1} \hat{d}_{1 \sigma}^{\dagger}\hat{d}_{1  \sigma}\!-\!\Delta_d \left( \hat{d}_{1 \downarrow} \hat{d}_{1 \uparrow} \!+\!\mbox{\rm h.c.} \right)\!, \ 
\end{eqnarray}
where $\Delta_d$ is dependent solely on the $QD_1-SC$ electrode coupling constant: $\Delta_d=\Gamma_S/2=\pi \sum_{k} |V_{k S}|^2 \delta(\omega-\xi_{k S})$.
In our work, we use the total coupling to conduction electrodes ($\Gamma_N$) as energy unit: $\Gamma_N=\Gamma_L + \Gamma_R= 2 \pi \sum_{k\beta} |V_{k,\beta}|^2 \delta(\omega-\xi_{k \beta})$.
Our time unit (t.u.) is given by the inverse of this energy scale $1\ \textrm{t.u.} = \hbar/\Gamma_{N}$. 
More detailed discussion on the time and energy scales is given in Section \emph{Quantified values of time}.

\section*{Methods}

Dynamics of coupled complex systems with discrete energy levels can be described in both time and energy scales. 
One of the scales is related to charge oscillations between subsystems. 
The period of these oscillations is usually associated with the energy of coupling between subsystems. 
In our case, apart from interdot coupling $t_{12\sigma}$, oscillations originating from the coupling of the system to superconducting electrode play an important role. 
All this oscillations do not thermalize unless the system is hybridized to continuum of states. 
Such continuum acts as damping force for classical oscillations \cite{taranko2018,baranski.baranska.21}. 
Therefore, if one is looking for a time of achieving steady-state solutions, the time scale is predominantly defined by energy of the coupling to metallic lead.
In our approach we first calculate charge oscillations on both quantum dots.

Using the Laplace transformation, we conduct analytic calculations of the exactly solvable model with the time-independent parameters, where coupling to continuum is assumed to be negligible, i.e.,  $\Gamma_N \rightarrow 0^{+}$. 
The main goal of the work, i.e.,  analysis of the above described double quantum dot system for arbitrary $\Gamma_N \neq 0$ and with the spin-independent interdot coupling ($t_{12\uparrow}(t) = t_{12\downarrow}(t) \equiv t_{12}(t) $), will be achieved by numerical calculations using the 4th order Runge-Kutta (4RK) method applied to equations of motion obtained in the Heisenberg representation.

\subsection*{Time dependent occupancies and charge oscillations}

In order to derive the time dependent occupation number $n_{i \sigma} (t) = \langle {\hat{d}}^{\dagger}_{i \sigma}(t){\hat{d}}_{i \sigma}(t)\rangle$, we start with equation of motion in Heisenberg representation for a given operator $\hat{O}$:
\begin{equation}
    \mathbf{i} \hbar \frac{d}{d t} \hat{O} (t) = [\hat{O}(t), \hat{H}].
\end{equation}
We calculate the set of such differential equations for creation ${\hat{d}}^{\dagger}_{1 \sigma}, {\hat{d}}^{\dagger}_{2 \sigma}, {\hat{c}}^{\dagger}_{k \sigma}$ and annihilation ${\hat{d}}_{1 \sigma}, {\hat{d}}_{2 \sigma}, {\hat{c}}_{k \sigma}$ operators. Next, for each fermion operator we conduct a Laplace transform ${\cal{L}}[\hat{O}(t)](s)$ defined as
\begin{equation}
    \hat{O}(s) = \int_0^\infty dt \exp{\left( -st \right) } \hat{O}(t)  \equiv {\cal{L}}[\hat{O}(t)](s),
\end{equation}
where $s$ is a complex variable. Such a procedure transforms a set of differential equations into a set of linear equations dependent on $s$ which are solvable analytically. We get analytical expression for each operator $\hat{O}(s)$. 
For example, in the case of time-independent $t_{12\sigma}$'s, for the annihilation operators ${\hat{d}}_{1 \uparrow}$ and ${\hat{d}}_{1 \downarrow}$, the $s$-dependent form looks as follows 
\begin{equation}
    {\hat{d}}_{1 \uparrow}(s) = - \frac{ \mathbf{i} \left[{\hat{A}}^{\dagger}_{\downarrow}(0)\Gamma_{S} + \mathbf{i} {\hat{A}}_{\uparrow}(0) M_{\downarrow}^{*}(s) \right]}{\Gamma_{S}^2 + M_{\downarrow}^{*}(s) M_{\uparrow}(s)}, \qquad
    {\hat{d}}_{1 \downarrow}(s) = - \frac{ \mathbf{i} \left[{\hat{A}}^{\dagger}_{\uparrow}(0)\Gamma_{S} + \mathbf{i} {\hat{A}}_{\downarrow}(0) M_{\uparrow}^{*}(s) \right]}{\Gamma_{S}^2 + M_{\uparrow}^{*}(s) M_{\downarrow}(s)},
\end{equation}
where
\begin{eqnarray}
 {\hat{A}}_{\sigma}(0) &=& -\mathbf{i} \sum_{k,\beta} \frac {V_{k \beta}} {s + \mathbf{i} \xi_{k \beta}} {\hat{c}}_{k \beta \sigma}(0) - \frac{ \mathbf{i} t_{12\sigma}}{s + \mathbf{i} \epsilon_2} {\hat{d}}_{2 \sigma}(0) + {\hat{d}}_{1 \sigma}(0) ,\nonumber \\
   M_{\sigma}(s) &=& s + \mathbf{i} \epsilon_1 + \sum_{k,\beta} \frac {|V_{k \beta}|^2} {s + \mathbf{i}\xi_{k\beta}} + \frac{t_{12\sigma}^2}{s + \mathbf{i} \epsilon_2} .
\end{eqnarray}
Note that equations above are obtained for a general case of spin-dependent $t_{12\sigma}$ ($t_{12\uparrow} \neq t_{12\downarrow}$) and arbitrary $\Gamma_{N}$, $\Gamma_{S}$, $\epsilon_1$, and $\epsilon_2$.

Now, we can get expressions for the occupation numbers in $s$-dependent form (i.e., expressions for $n_{i \sigma} (s) = \langle {\hat{d}}^{\dagger}_{i \sigma}(s){\hat{d}}_{i \sigma}(s)\rangle$) in the limit of $\Gamma_N \rightarrow 0$. 
The general equations are rather complex, therefore, we present here the results obtained in the conditions: $\epsilon_1=\epsilon_2=0$ and $n_{2 \sigma}(0)=0$.
After removing pairs of operators that give zero and applying the inverse Laplace transform for a given operator ${\cal{L}}^{-1}[\hat{O}(s)](t)$, for the spin-independent inter dot coupling, i.e., $t_{12\uparrow} = t_{12\downarrow}=t_{12}$, we formulate the time dependence of expectation values $n_{1 \sigma}(t)$ and $n_{2 \sigma}(t)$:
\begin{eqnarray}
\label{n1odt} 
    n_{1 \sigma} (t) & = & - [1 - {n}_{1 \bar{\sigma}}(0)] L_1^2  + {n}_{1 \sigma}(0) (L_2 + L_3)^2 ,\\
\label{n2odt}
    n_{2 \sigma} (t) & = & [1 - {n}_{1 \bar{\sigma}}(0)] L_4^2 - {n}_{1 \sigma}(0) (L_5 + L_6)^2 ,
\end{eqnarray}
where $n_{i \sigma} (0) = \langle \hat{d}^{\dagger}_{i \sigma}(0)\hat{d}_{i \sigma}(0)\rangle$ and
\begin{eqnarray}
    \label{eq:laplace}
    L_1 & = & {\cal{L}}^{-1} \left[ \frac{\mathbf{i}  \Gamma_s}{\Gamma_s^2 + \left( s + \frac{t_{12}^2}{s} \right)^2} \right], \qquad \ 
    L_2 = {\cal{L}}^{-1} \left [\frac{s}{\Gamma_s^2 + \left(s + \frac{t_{12}^2}{s}\right)^2} \right ], \quad   
    L_3  =  {\cal{L}}^{-1} \left [\frac{t_{12}^2}{s \left[\Gamma_s^2 + \left(s + \frac{t_{12}^2}{s} \right)^2\right]} \right],  \qquad \\ 
    L_4  & =  & {\cal{L}}^{-1} \left [\frac{- t_{12} \Gamma_s}{s \left[\Gamma_s^2 + \left(s + \frac{t_{12}^2}{s} \right)^2 \right]} \right], \quad
    L_5  =  {\cal{L}}^{-1} \left [\frac{\mathbf{i}  t_{12}}{\Gamma_s^2 + \left(s + \frac{t_{12}^2}{s} \right)^2} \right ],   \quad
    L_6 = {\cal{L}}^{-1} \left [\frac{ \mathbf{i} t_{12}^3}{s^2 \left[\Gamma_s^2 + \left(s + \frac{t_{12}^2}{s} \right)^2 \right]} \right]. \nonumber
\end{eqnarray}
%

\begin{figure}[t]
        \centering
	\includegraphics[width=\wymiarlargetwo]{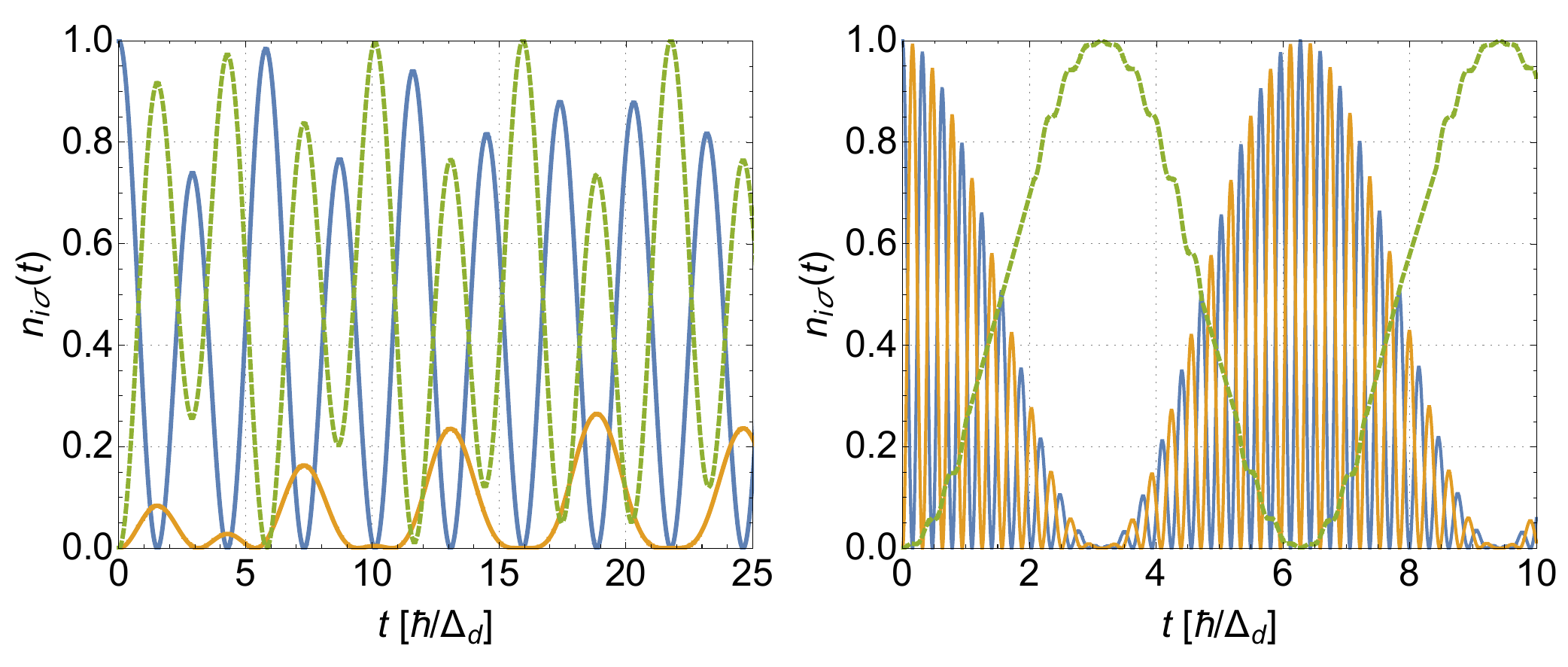}
	\caption{\textbf{Charge oscillations in the limit of $\Gamma_N \rightarrow 0$.}
        Occupation numbers  $n_{1 \sigma}(t)$ (blue lines) and $n_{2 \sigma}(t)$ (orange lines) obtained for $t_{12}=0.3\Delta_d$ (left) and $t_{12}=10\Delta_d$ (right) with the initial conditions: $n_{1 \sigma}(0)=1$ and $n_{2 \sigma}(0)=0$ [cf. equations (\ref{n00}) and (\ref{n02})]. 
        Green dashed lines indicate charge transferred (per spin) into the region of the SC electrode.}
	\label{fign}
\end{figure}

Here, we will find the explicit equations for the time-dependence of occupation number $n_{i \sigma}(t) = \langle \hat{d}^\dagger_{i \sigma}(t) \hat{d}_{i \sigma} (t) \rangle$ in the limit of $\Gamma_N \rightarrow 0$ and $\epsilon_1=\epsilon_2=0$, time-independent $t_{12\sigma} \equiv t_{12} $ for $\sigma=\uparrow,\downarrow$.
We also assume that at the initial moment ($t=0$), the side attached quantum dot is empty, i.e., $n_{2 \sigma}(0)=0$ (for both $\sigma$'s). 
Note that, in such a case, time dependence of $n_{i\sigma}$ is spin-independent ($n_{i\uparrow}(t)=n_{i\downarrow}(t)$).
Taking into account this simplifications, we obtain from (\ref{n1odt}) and (\ref{eq:laplace}) (and $\Delta_d=\Gamma_S/2$)
\begin{eqnarray}
n_{1 \sigma}(t)&=&
\frac{n_{1 \sigma}(0)\left[\omega_{+}\cos(\omega_{+} t) + \omega_{-}\cos(\omega_{-} t)\right]^2}{\Delta_d^2+4t_{12}^2}  
+ \frac{h_{1 \bar{\sigma}}(0) \left[\omega_{+}\sin (\omega_{+} t)-\omega_{-}\sin (\omega_{-} t)\right]^2}{\Delta_d^2+4t_{12}^2},
\label{n00}
\end{eqnarray}
where $h_{1 \sigma}(0)=1-n_{1 \sigma}(0)$ and $\bar{\sigma}$ denotes spin direction opposite to $\sigma$. 
Charge oscillations are defined by two frequencies 
\begin{equation*}
\omega_{\pm} = \sqrt{\tfrac{1}{2}\left(\Delta_d^2+2t_{12}^2 \pm \Delta_d\sqrt{\Delta_d^2+4t_{12}^2} \right)}.
\end{equation*}
We note that, for the vanishing interdot coupling ($t_{12}\rightarrow 0$),  frequency $\omega_{-}$ is equal to $0$, while $\omega_{+} \rightarrow \Delta_d$. 
This brings the equation (\ref{n00}) to the form of $n_{1 \sigma}(t)=n_{1 \sigma}(0)\cos^2(\Delta_d t)+h_{1 \bar{\sigma}}(0)\sin^2(\Delta_d t)$, which reproduces the result obtained for a single QD placed between metallic and superconducting leads  for vanishing $\Gamma_N$ (see Ref. \cite{taranko2018}). 
In the case of the most interest to us, when the interdot coupling is significantly smaller then the effective pairing (but not negligible), the resultant oscillations have low- and high-frequency modes. 
High-frequency mode is predominantly governed by the coupling of the dot to the superconducting reservoir $\omega_{+} \sim \Delta_d$. 
Low frequency ($\omega_{-}$), in turn, is related to the interdot coupling $t_{12}$. 
It is also interesting to investigate the charge fluctuations that occur on $QD_2$. 
Oscillations of $n_{1 \sigma}(t)$ are associated to a charge flow from $QD_1$ to $QD_2$ (and vice versa) and a charge flow between $QD_1$ and the SC electrode. 
In our model, $QD_2$ is tunnel coupled only to $QD_1$, therefore changes of $n_{2 \sigma}(t)$ originates solely from the interdot charge flow. 
Considering the same conditions as previously (i.e., $\epsilon_{1}=\epsilon_{2}=0$, and $n_{2 \sigma}(0)=0$), occupation  $n_{2 \sigma}(t)$ can be calculated  from (\ref{n2odt}) by performing the inverse Laplace transformation of the equations (\ref{eq:laplace}) given previously. One obtains
\begin{eqnarray}
n_{2 \sigma}(t) &\simeq& h_{1 \bar{\sigma}}(0) t_{12}^2\frac{\left[\cos\left(\omega_{+} t\right) - \cos\left(\omega_{-} t \right)\right]^2}{\Delta_d^2+4t_{12}^2}  
+ n_{1 \sigma}(0)t_{12}^2\frac{\left[\sin (\omega_{+} t)-\sin (\omega_{-} t)\right]^2}{\Delta_d^2+4t_{12}^2}. \ 
\label{n02}
\end{eqnarray}
The occupancies of both dots and the dot-lead charge flow (per spin) calculated from equations (\ref{n00}) and (\ref{n02}) are presented in Figure \ref{fign}.
Assuming that at the initial conditions, $QD_1$ was full and $QD_2$ was empty (i.e., $n_{1 \sigma}(0)=n_{1 \bar{\sigma}}(0)=1$, $n_{2 \sigma}(0)=n_{2 \bar{\sigma}}(0)=0$), the charge transferred to the SC electrode can be expressed by deficiency of charge in both QD's, i.e., $1-n_{1 \sigma}(t)-n_{2 \sigma}(t)$.

For weak interdot coupling (i.e., $t_{12} < \Delta_d$), the amplitude of charge oscillations on $QD_{2}$ is strongly reduced (compared to the amplitude of $n_1(t)$), however, frequencies of oscillations for both quantum dots are very close (cf. the left panel of Figure \ref{fign}). 
As the occupation number of $QD_1$ oscillates in a full range (from $0$ to $1$) while the amplitude of oscillations in $QD_2$ is significantly reduced, we conclude that only part of the charge flowing from SC reservoir to $QD_1$ is transmitted to $QD_2$.

Although in our work we mainly focus on the limit of the weak (but not negligible) interdot coupling, it is very interesting to investigate charge oscillations for $t_{12} \gg \Delta_d$  (cf. the right panel of Figure \ref{fign}). 
In this so-called molecular regime, two quantum dots form a single molecule placed on the top of the superconductor. 
Consequently, charge oscillates are very fast between dots (cf. oscillations of blue and orange line in the figure), while the weak coupling of the molecule to the SC reservoir causes slow charge flow from the dots to SC electrode and back to the double-quantum-dot (DQD) molecule (cf. green line on the figure).
Charge flow from the DQD to the SC is possible only if a charge on  $QD_1$ dot is nonzero.
Conversely, charge flow from the SC to the DQD is possible only if the $QD_1$ is at least partially empty. 
Therefore, in the case of $t_{12}\gg\Delta_d$ (where charge between dots is swapped rapidly), the charge transfer to and from the SC electrode appears sequentially after each short-term cycle between the dots.
Such behavior can be inspected in Figure \ref{fign} (right), where charge transfer between the DQD molecule and the SC (marked by the green line) occurs sequentially after each interdot charge transfer.
In discussed here limit (i.e., $t_{12} \gg \Delta_d$), frequencies $\omega_{+}$ and $\omega_{-}$ become very close to each other ($\omega_{+} \approx \omega_{-}$). 
Introducing $\frac{1}{2}(\omega_{+}+\omega_{-}) \equiv \omega_m$ and $\frac{1}{2}(\omega_{+}-\omega_{-}) \equiv \KK{\delta_\omega}$ (with $\KK{\delta_\omega} \ll \omega_m$), we obtain, in this limit, the equations for charge oscillations on both dots in the form of a square of classic beats
$n_{1 \sigma}(t)\simeq n_{1 \sigma}(0)\left[\cos(\omega_{m} t)\cos(\delta_{\omega} t)\right]^2+ h_{1 \bar{\sigma}}(0) \left[\KK{\cos} (\omega_{m}  t)\sin (\delta_{\omega} t)\right]^2$,
$n_{2 \sigma}(t)\simeq n_{1 \sigma}(0)\left[\cos(\omega_{m} t)\sin(\delta_{\omega} t)\right]^2+ h_{1 \bar{\sigma}}(0) \left[\sin (\omega_{m}  t)\sin (\delta_{\omega} t)\right]^2$. 
High-frequency beats (given by $\omega_m$) \KK{refer} to interdot oscillations, while the low-frequency mode (given by $\delta_\omega$) is related to a charge transfer between the molecule and the SC electrode. 
We note that, at the limit of $t_{12} \rightarrow \infty$, low-frequency mode $\delta_{\omega}$ is equal to $\Delta_d/2$, which reproduces the case of a single molecule placed on top of superconductor, but with the twice weaker coupling  \cite{taranko2018}.

\subsection*{Numerical approach}

Asymmetric line-shapes being a symptom of electron scattering on a side level can be practically investigated by inspection of time-dependent charge current induced by a bias voltage.
Our aim is to calculate the charge current $I_{L}(t,eV)$ flowing through one of  the metallic electrodes upon bias voltage $V$  applied to another one ($e$ denotes the elementary charge), i.e., the source-drain voltage between the normal electrodes. 
We assume that chemical potential of the SC electrode and the metallic $R$ electrode are equal and we measure the energy with respect to these potentials ($\mu_{SC}=\mu_{R}=0$).
The pronounced resonant features can be inspected in differential conductivity $G(t,eV)=\frac{d }{d V}I(t,eV)$. 
The charge flowing through $L$ electrode is given by average change of the electron number in the $L$ lead
\begin{eqnarray}
I_{L \sigma} (t, V)= -e \left \langle \frac{d N_{L \sigma}}{dt} \right \rangle ,
\label{it}
\end{eqnarray}
where $N_{L \sigma} = \Sigma_k \hat{c}^{\dagger}_{k L \sigma} \hat{c}_{k L \sigma}$.

In order to find time-dependent statistical averages we derive the closed set of ordinary differential equations of motion in terms of the Heisenberg notation $\frac{d}{dt}\hat{O}(t)=\frac{\mathbf{i}}{\hbar}[\hat{H},\hat{O}]$ and apply the 4th order Runge-Kutta (RK4) method for numerical calculations of time evolution for each average. 
Details of this procedure were introduced in the previous work on dynamics of Majorana-QD hybrid~\cite{baranski.baranska.21}.
The equations of motion in the Heisenberg representation have been derived using \textsc{SNEG} library created by R.~\v{Z}itko~\cite{Zitko_2011}.

\subsubsection*{Quench protocol}

We assume that initially ($t \leq 0$) all parts of the system (both QD's and three electrodes) are separated. 
Such initial condition can be met by setting the averages comprising operators referring to different parts of the system like $\langle \hat{d}^{(\dagger)}_{i \sigma}(0) \hat{c}^{(\dagger)}_{k \beta \sigma '}(0) \rangle$, $\langle \hat{d}^{(\dagger)}_{i \sigma}(0) \hat{d}^{(\dagger)}_{j \sigma'}(0) \rangle$ to be equal $0$. 
The assumption that the quantum dots are separated from the SC electrode for $t<0$ also imposes that there was no pairing potential in the region of the QDs. 
This requirement is met by $\langle \hat{d}^{\dagger}_{i \sigma}(0) \hat{d}^{\dagger}_{i \bar{\sigma}}(0) \rangle = \langle \hat{d}_{i \bar\sigma}(0) \hat{d}_{i \bar{\sigma}}(0) \rangle=0$. 
The average numbers of electrons in $L$ and $R$ leads are given by the Fermi distribution function $\langle \hat{c}^{\dagger}_{k \beta \sigma}(0) \hat{c}_{k \beta \sigma}(0) \rangle = \lbrace 1 + \exp\left[ (\xi_{k \beta}-\mu_{\beta}) / (k_B T) \right] \rbrace^{-1}$.
In order to do not confuse the effect of formation of quasiparticles with interference effects we developed two step procedure. 
First, we assume that at time $t=0$ $QD_{1}$ is connected only to the external electrodes (i.e., $L$, $R$ and $S$) keeping $t_{12}=0$. 
Static results for single QD in such heterostructure show that evolution should lead to the formation of two quasiparticle Andreev states located near $\pm\sqrt{\epsilon_{1}^{2} +\Delta_{d}^{2} }$ (cf., e.g., Refs. \cite{Baraski2013,Bauer2007}).
When these states achieve its static values and time fluctuations will vanish (let say at time $t=t_0$), we abruptly connect the second quantum dot  [with constant interdot coupling term, $t_{12\sigma}(t) \equiv t_{12}(t)=t^{0}_{12} \theta( t-t_0 ) $ for $\sigma = \uparrow,\downarrow$]. 
From this moment, scattering of electrons on the side level becomes possible and we can observe the evolution of interference patterns emerging for voltages ($eV$) close to $ \pm \epsilon_2$. 
In all our calculations we assume the quench time to be $t_0=20 \ \hbar/\Gamma_N$, which, in the considered energy scale, is much larger than the relaxation time after connecting the dot to the external electrodes.

\begin{figure}[t]
        \centering
	\includegraphics[width=\wymiarmediumtwo]{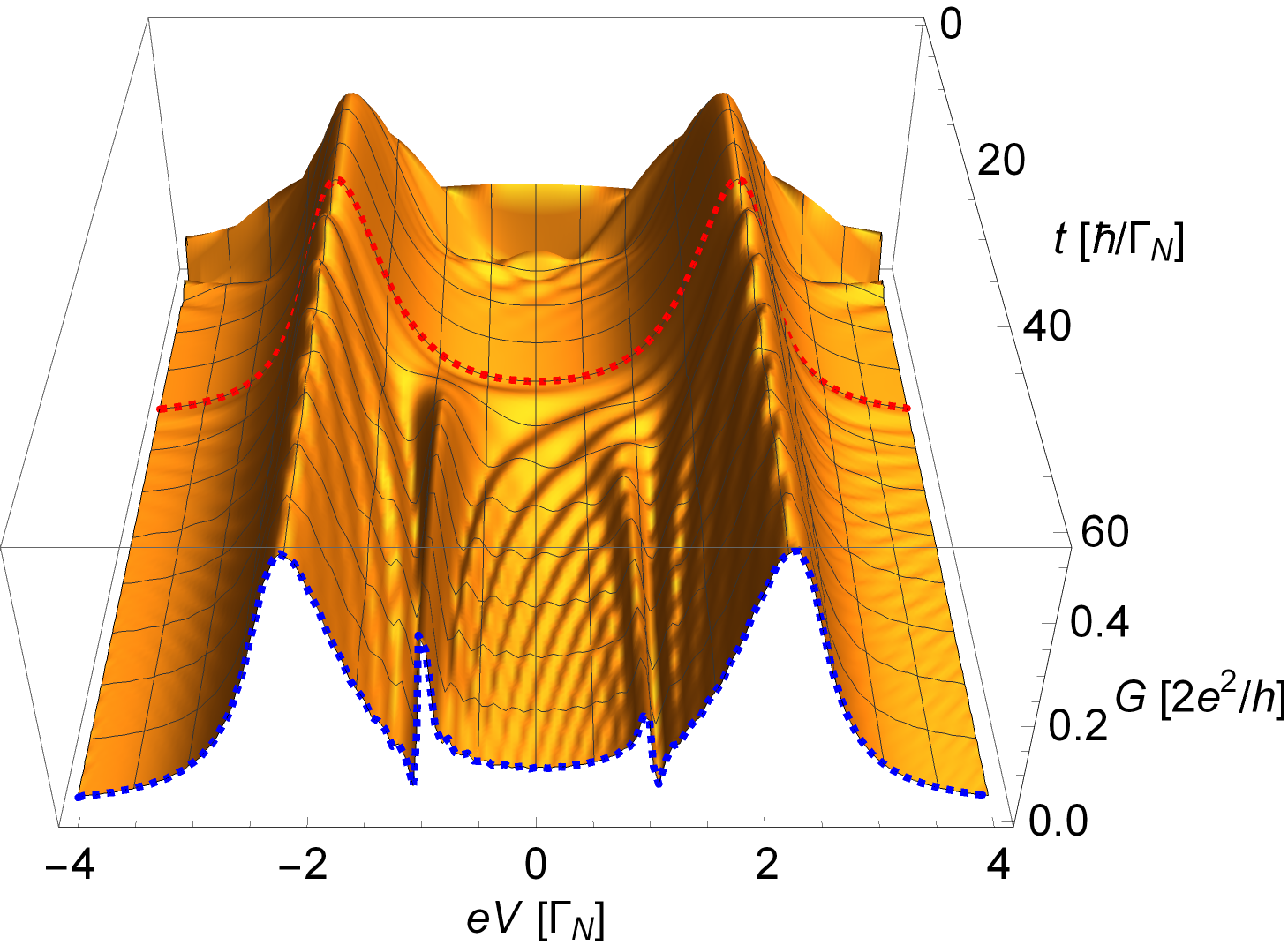}
	\caption{\textbf{Differential conductivity $G$ $[2e^2/h]$ as 
        a function of bias voltage $eV$ $[\Gamma_{N}]$ and time $t$ $[\hbar/\Gamma_{N}]$.}
        The results are obtained for $\Delta_{d}=2\Gamma_N$, $\epsilon_{2}=\Gamma_{N}$ and $t^{0}_{12}=0.3{\Gamma_{N}}$. 
	The red dashed line indicates the moment ($t_{0}=20$ $\hbar/\Gamma_{N}$) at which tunneling between dots is turned on. 
	The blue dashed line shows the temporary shape of the evolving resonant states 
        at $t=60$ $\hbar/\Gamma_{N}$. 
	}
	\label{fig2c}
\end{figure}

\subsubsection*{Static calculations}

In order to verify obtained results we can calculate also the static (time-independent) conductivity. 
Conductivity calculated using the RK4 method for ($t \rightarrow \infty$) should reproduce these results. 
In three terminal heterostructure comprising two metallic and one superconducting electrode, charge transport is provided by three types of processes: 
(i) ballistic single electron transfer (ET) from $L$ to $R$ electrode,
(ii) direct Andreev reflection (DAR), where single electron from $L$ electrode is converted to a Cooper pair propagating in the SC electrode with simultaneous reflection of hole back to $L$ electrode, and
(iii) cross Andreev reflection (CAR), in which the hole is reflected to second metallic electrode ($R$). 
For the static case these three processes can be evaluated using the following Landauer-like formulas \cite{Michalek2013,Michalek2017} 
\begin{eqnarray}
J^{ET}_{L}=\frac{2e}{h} \int d \omega \Gamma_{L} \Gamma_{R} |G^{r}_{11}|^2 (f_{L} - f_{R}),
\label{st1}
\end{eqnarray}
\begin{eqnarray}
J^{DAR}_{L}=\frac{2e}{h} \int d \omega \Gamma_L^{2} |G^{r}_{12}|^{2} (f_{L} - \tilde{f}_{L}),
\label{st2}
\end{eqnarray}
\begin{eqnarray}
J^{CAR}_{L}=\frac{2e}{h} \int d \omega \Gamma_{L} \Gamma_{R} |G^{r}_{12}|^{2} (f_{L} - \tilde{f}_{R}),
\label{st3}
\end{eqnarray}
where $G^r_{ij}$ are matrix elements of the retarded Green functions of $QD_1$ in the Nambu representation (given, e.g., in Ref. \cite{baranski.zienkiewicz.20}), 
$f_{\beta} =\left\{1+\exp{\left[(\omega-\mu_{\beta})/(k_B T)\right]}\right\}^{-1}$ and 
$\tilde{f}_{\beta}=\left\{ 1+\exp{ \left[ (\omega+\mu_{\beta})/(k_B T) \right] } \right\}^{-1}$ are the Fermi distributions of electrons and holes, respectively.
In our calculations we assume that voltage ($V$) is applied to $L$ electrode while the chemical potential of $S$ and $R$ electrodes are equal and energies are measured with respect to them ($\mu_{R}=\mu_{S}=0$). 
Time-dependent current calculated using the RK4 method accounts for all \KK{these} three processes together.

\section*{Results and discussion}

\begin{figure}[t]
\centering
	\includegraphics[width=\wymiarlarge]{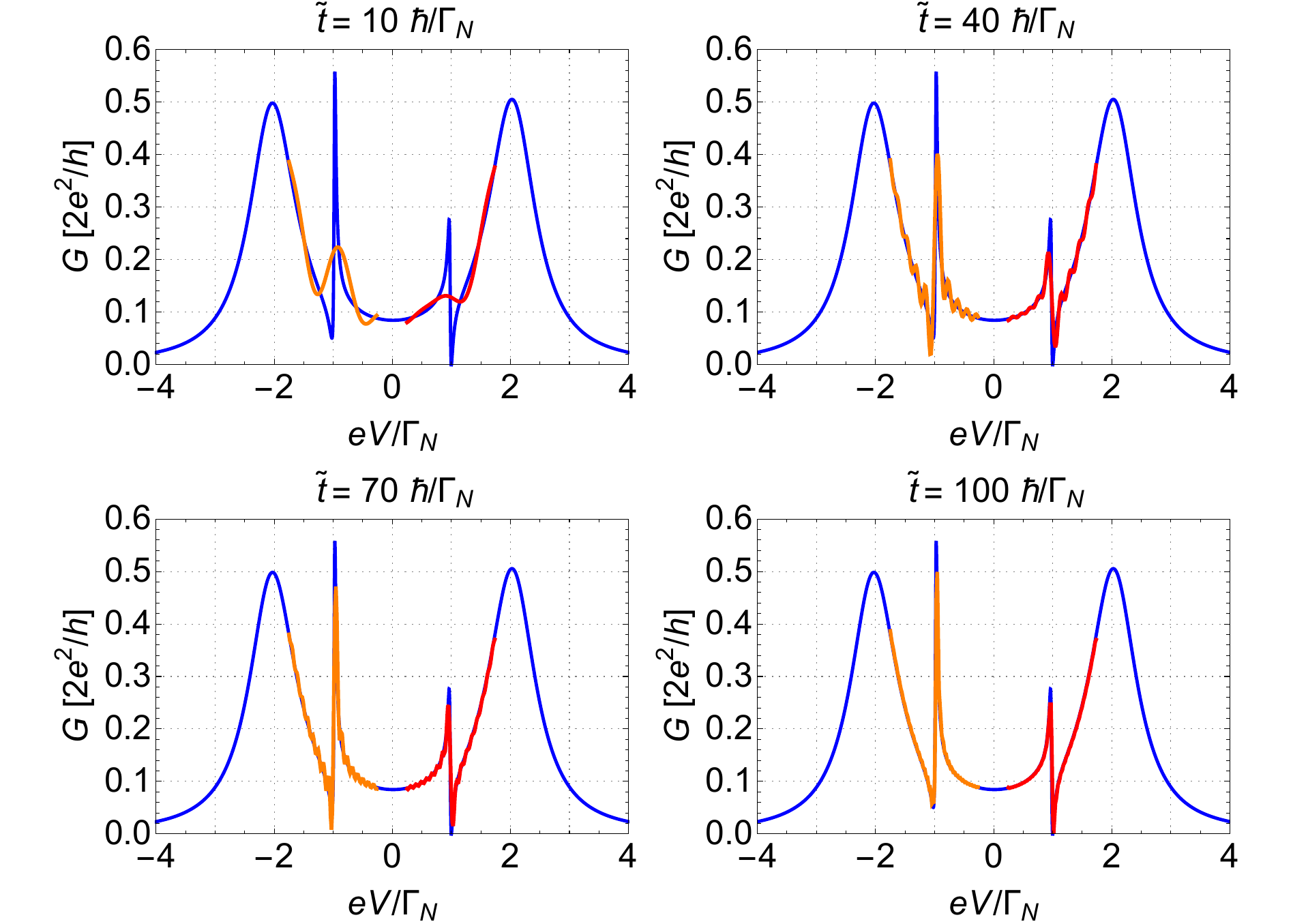}
	\caption{\textbf{The comparison of the resonant features near $\epsilon_{2}$ (red lines) and 
            $-\epsilon_{2}$ (orange lines) obtained for several different times ($\tilde{t} =t-t_{0}$) 
            with the static ($t \rightarrow \infty$) conductance (the blue lines).} 
        The results are presented for $\tilde{t}= 10, \ 40,\ 70,\ 130$ $[\hbar/\Gamma_N]$, respectively (as labelled). 
        $V$ denotes the bias voltage.
	Other parameters used here are the same as in Figure \ref{fig2c}.}
	\label{fig2b}
\end{figure}

In Figure \ref{fig2c} we present a time evolution of the total conductance $G$ (in $2e^2/h$ units) versus applied bias (source-drain) voltage $eV$ (in units of $\Gamma_{N}$). 
In accordance to the assumed quench protocol (cf. Section \emph{Quench protocol}), at the beginning $QD_{1}$ is connected only to the external electrodes and two Andreev states build up over time. 
We noticed that, after approximately $10$--$15$ $\hbar/\Gamma_{N}$, these states saturate and all fluctuations are suppressed. 
To be sure that the process of building up of the Andreev states does not affect formation of the interference features, we set the quench time safely later, i.e., at $t_{0}=20$ $\hbar/\Gamma_{N}$. 
At this moment, the interdot connection is established with coupling strength $t^{0}_{12}=0.3\Gamma_{N}$ and the resonant characteristics start to evolve. 
The moment of abrupt establishing of the interdot coupling is underlined by the red dashed line in Figure \ref{fig2c}. 
The blue dashed line highlights the shape of both resonances obtained at $t=60$ $\hbar/\Gamma_{N}$.

We note that a well-pronounced AF feature (i.e., this near $-\epsilon_{2}$) starts to develop almost instantaneously after the abrupt connection is established, while only small fluctuations appear near $ eV \simeq \epsilon_{2}$. 
It is surprising that even though the direct scattering feature is not developed yet, the feature announcing its superconducting response is building up so vigorously. 
The non-equilibrium response for scattering indicates that whenever electron being a component of a local pair is involved in a given process even by a small fluctuations, the second spin component reacts instantaneously and robustly.
Considering time required for achieving equilibrium conditions, however, one need to have in mind that, in the static solutions (i.e., for $t \rightarrow \infty$), the maximum of the feature announcing the SC response for scattering, i.e., the AF feature, is magnitude of order higher than the maximum of the ordinary Fano resonance feature. 
Measures of development for each resonance should be thus considered relative to its final amplitude rather than one to another.

In this regard, in Figure \ref{fig2b}, we overlay the resonant features obtained for several time parameters on plots of static ($t \rightarrow \infty$) conductance obtained using equations (\ref{st1})--(\ref{st3}). 
We note that, after $\tilde{t}=10\ \hbar/\Gamma_{N}$  from the quenching of QD$_2$, the feature announcing direct resonance (red line) is given only by tiny fluctuations, while for the indirect resonance, a well-pronounced peak is already visible. 
However, after $\tilde{t}=40\ \hbar/\Gamma_{N}$,  development of both resonances (relative to their static amplitude) appears to be comparable. 
At $\tilde{t}=70\ \hbar/\Gamma_{N}$, the Fano-like resonance is close to the stable solution and only small adjustment of its shape is contributed between $\tilde{t}=70\ \hbar/\Gamma_{N}$ and $\tilde{t}=130\  \hbar/\Gamma_{N}$. 
In contrary, the amplitude of the AF resonance is still building up even after $\tilde{t}=130\ \hbar/\Gamma_{N}$. 
From the picture above, it seems that although short after quench the AF resonance builds up very fast (comparing to the ordinary Fano feature), the stable solution for the ordinary Fano resonance is achieved slightly faster than for the AF resonance. 
To deliver quantified data for development of both these resonances, we will inspect closely the time required for development of a stable peak for each resonance.

\begin{figure}[t!]
    \centering
    \includegraphics[width=\wymiarmediumone]{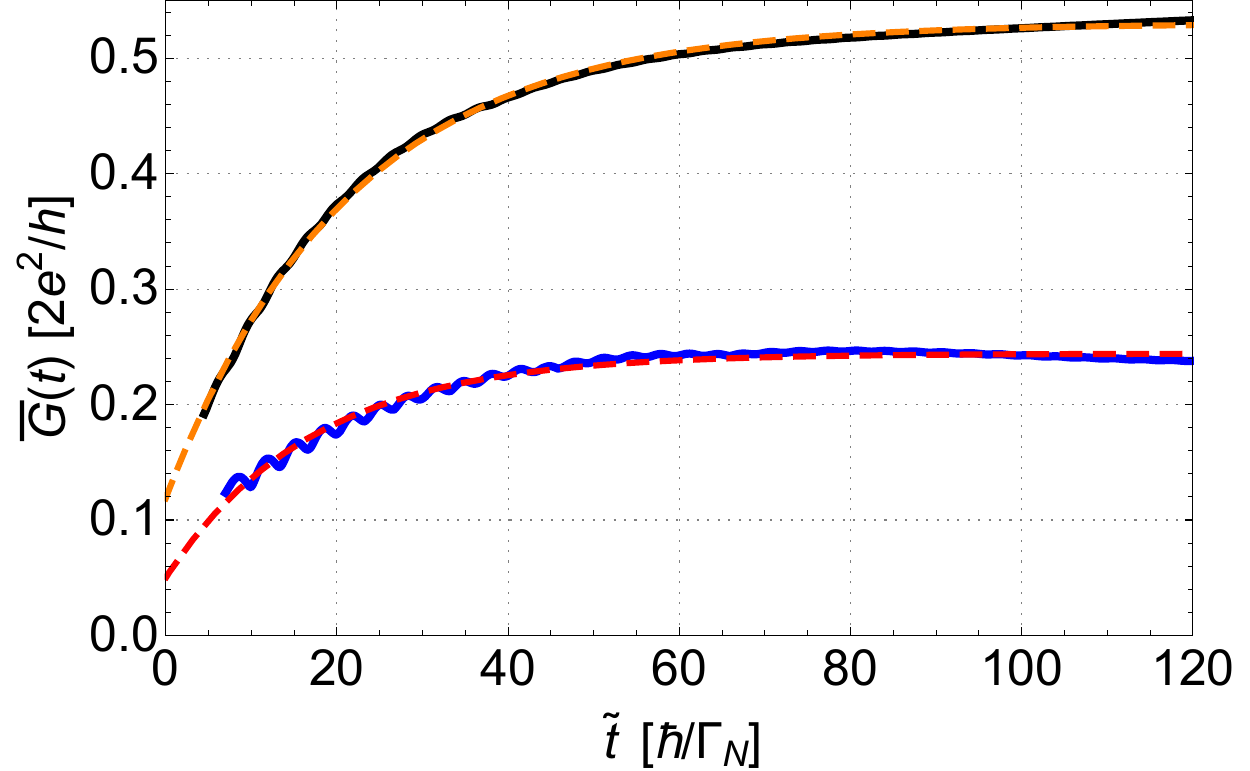}
    \caption{%
	\textbf{Amplitudes of the local maxima near the ordinary Fano resonance (blue solid line) 
        and the anomalous Fano feature (black solid line) as a function of time $\tilde{t}=t-t_0$ [$\hbar/\Gamma_{N}$].}
        The corresponding fitted exponential curves are shown by dashed red and orange lines. 
        Results obtained for $t^{0}_{12}=0.4\Gamma_{N}$. 
	Other parameters are the same as in Figure \ref{fig2c}.
    } 
    \label{AFfit}
\end{figure}

\section*{Quantified analysis of time evolution of the resonant features}

\begin{table}[b!]
    \caption{Relaxation factors $\tau_F$ and $\tau_{AF}$ obtained for several couplings $t^{0}_{12}$ and corresponding equilibrium times $t^{F}_{eq}$, $t^{AF}_{eq}$.}
    \label{tab1}
  \begin{center}
    \begin{tabular}{c|c|c|c|c}
      $t^{0}_{12}$ $[\Gamma_{N}]$ & $\tau_{F}$ $[\hbar/\Gamma_N] $ & $\tau_{AF}$ $[\hbar/\Gamma_N] $ & $t^{F}_{eq}$ [ns] &
      $t^{AF}_{eq}$ [ns]
      \\
      \hline
 0.3 &  23.24 & 29.35 & 8.86 & 10.97\\
 0.4 &  17.06 & 21.24 & 6.38 & \ 7.94\\
 0.5 &  12.79 & 14.75 & 4.78 & \ 5.52\\
 0.6 &  10.02 & 11.12 & 3.74 & \ 4.16 \\
    \end{tabular}
  \end{center}
\end{table}

\begin{figure}[t!]
        \centering
	\includegraphics[width=\wymiarmediumone]{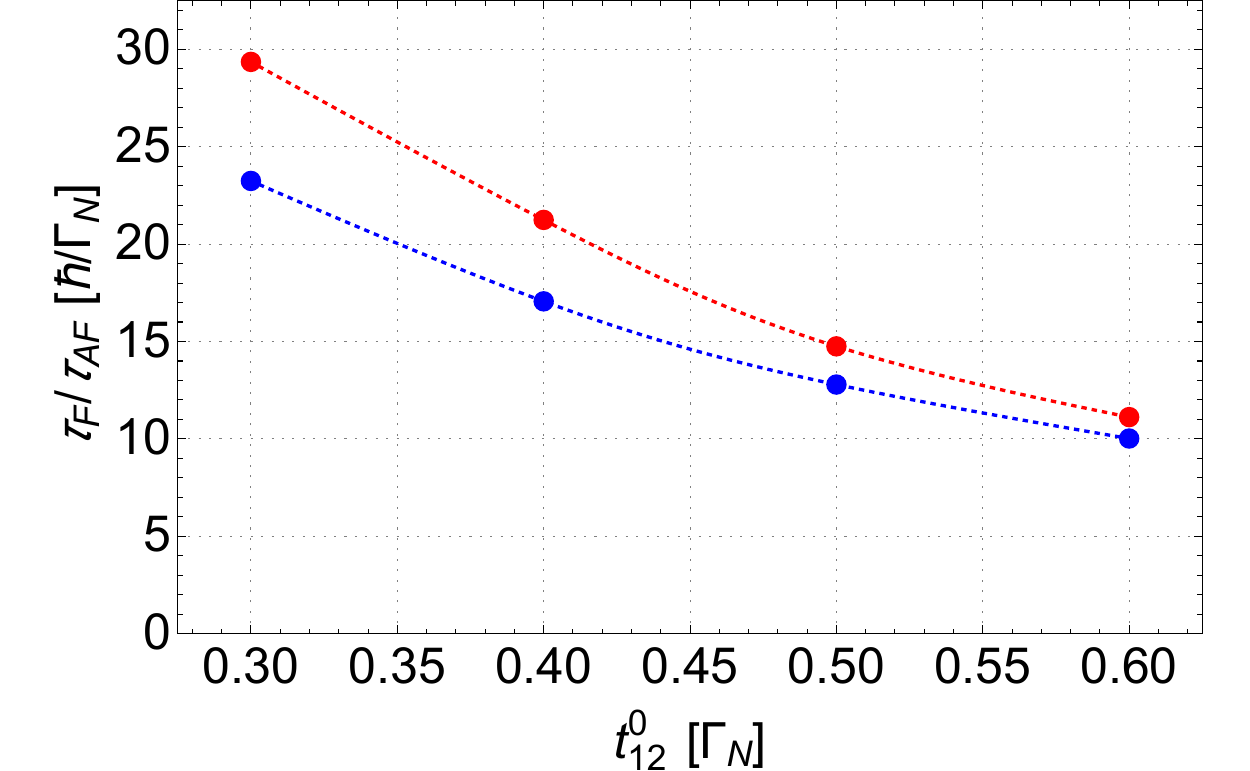}
	\caption{\textbf{Dependencies of the equilibrium times on the interdot coupling.}
            Times $\tau_F$ and $\tau_{AF}$ (in the units of $\hbar/\Gamma_{N}$)  as a function of the interdot coupling $t^{0}_{12}$ obtained for 
            the ordinary Fano resonance (blue line) and the anomalous Fano feature (red line), respectively.}
	\label{taudepence}
\end{figure}

We noticed that evolution of the OF and the AF peaks in time resembles the exponential growth (cf. Figure \ref{AFfit} for $t^{0}_{12}=0.4\Gamma_{N}$). 
In order to quantify the time required for formation of the stable resonant features, we fit the exponential function describing decay of a difference between the actual and the final value of maxima. 
This function is defined as
\begin{eqnarray}
\bar{G}(t) = \bar{G}(\infty) -[\bar{G}(\infty)- \bar{G}(t_0)] \exp{ \left[ (t_0-t )/ \tau \right] } ,
\label{function}
\end{eqnarray}
where $\bar{G}(\infty)$ is the peak amplitude in the static conditions (i.e., at $t\rightarrow + \infty$), $\bar{G}(t_0)$ is a value of the local maxima at the initial moment of the quench (at $t_0$), and the fitted parameter $\tau$ represents the interval, in which a mismatch between the initial conductivity and
the equilibrium conductance diminishes $e \approx 2.71$ times.
The calculations have been performed for a few amplitudes of the interdot coupling strengths ranging from $t^{0}_{12}=0.3\Gamma_{N}$ to \KK{$t^{0}_{12}=0.6 \Gamma_{N}$}.
We find that the characteristic time parameter $\tau_{AF}$ obtained for the anomalous Fano is 
larger than its counterpart $\tau_{F}$ obtained for the ordinary Fano resonance. 
The difference $\tau_{AF}-\tau_{F}$ is larger for weak interdot coupling $t^0_{12}$. 
Data of obtained $\tau$ for both resonances are collected in the second and third column of Table~\ref{tab1}. 
The dependencies of both $\tau$'s as a function of the interdot coupling strength are presented in Figure \ref{taudepence}. 

Time of achieving the equilibrium conditions $t_{eq}$ could be considered as time after which the difference between the initial and the final amplitudes of a given resonant feature decrease by $95\%$. 
In order to deliver information on a typical timescale of the resonances in tangible units (i.e., nanoseconds), in the table, we present also $t_{eq}$ assuming that the dot-lead coupling $\Gamma_{N}$ is equal to 50 $\mu$eV. 
The energy scale of this magnitude is typical for experiments considering \KK{nanoscale} objects coupled with superconductors \cite{Lee_2013,Junger_2019}.

\begin{figure}[t!]
    \centering
    \includegraphics[width=\wymiarlargetwo]{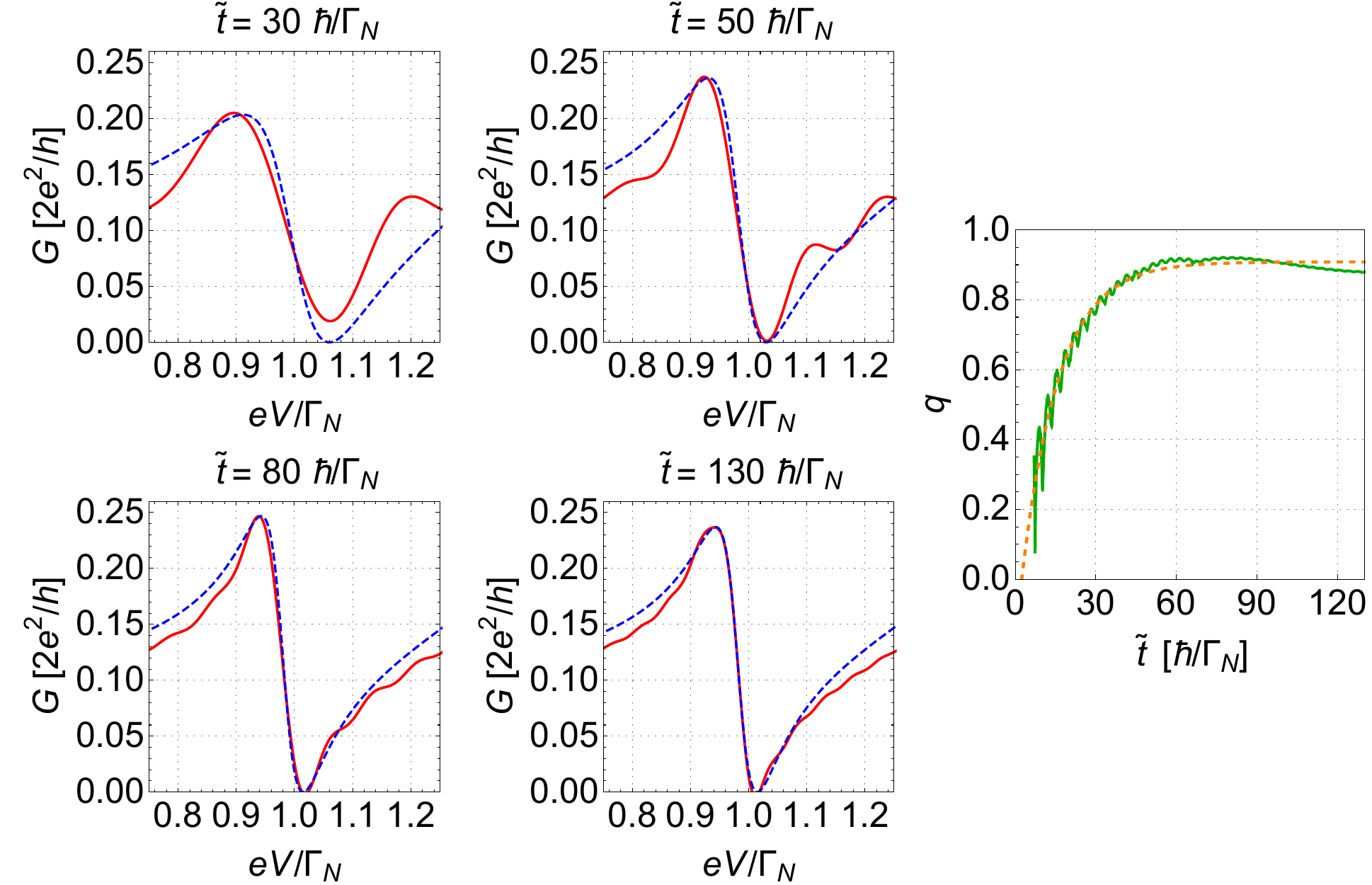}
    \caption{%
	\textbf{The evolution of the ordinary Fano resonant feature in time.}
        The four panels on the left \KK{show} the comparison of the resonant feature (solid red lines) with the best fit of the Fano shape curve (blue dashed lines) obtained for several different times ($\tilde{t} =t-t_{0}$, as \KK{labeled}).
        The right panel presents the time evolution of the parameter $q$ of the fitted Fano curve to the data (solid green line). 
        The orange dashed line shows the corresponding fitted exponential curve with $\tau_q$ parameter. 
        Results obtained for $t^{0}_{12}=0.4\Gamma_N$. Other parameters are the same as in Figure \ref{fig2c}.
    } 
    \label{FanoFitting}
\end{figure}

The Fano-type interference peaks are characterized by asymmetric line-shapes originating from the close coexistence of resonant enhancement and resonant suppression of transmission. The function resembling the Fano-like shapes reads as
\begin{eqnarray}
 F(\omega)=\alpha(\omega) \frac{(q \Gamma_{K}/2+\omega-\omega_{res})^2}{(\Gamma_{K}/2)^2+(\omega-\omega_{res})^2} 
 \label{fano}
\end{eqnarray}
where $\alpha(\omega)$ is the background transmission (which, in our case, is the undisturbed Andreev feature and can be obtained from equations (\ref{st1})--(\ref{st3}) at $t=t_0$, i.e., before the quench), $\omega_{res}$ is the energy of the resonant (quasi-discrete) level, 
$\Gamma_K$ is the broadening of the resonant level and $q$ is the so-called asymmetry parameter (cf. also Ref. \cite{baranski.zienkiewicz.20} as well as original works of the resonance on a flat background\cite{Fano1935,Fano1961}). 
One of the crucial parameters describing the Fano shape is an asymmetry parameter $q$. 
If $q=0$, the Fano function resembles symmetric deep, whereas, for $q \rightarrow \infty$, the Fano function develops into the Breit-Wigner (Lorentz) distribution. 
It is interesting to investigate the time evolution of the asymmetry parameter $q$. 
In order to study time dependence of $q$ parameter, the Fano function (\ref{fano}) is fitted to the features evolving around $eV=\epsilon_{2}$ after the quench. 
From such a fit, the values of $q$ are extracted.
For this purpose, we use the procedure developed and described in earlier work \cite{baranski.zienkiewicz.20}. 
Such fitting makes sense if the original shape of the transmission  has well-pronounced both local minima and maxima. 
At the moment of the quench, the shape of conductivity near the resonant level (i.e., at $eV\approx \Gamma_{N}$) is given by a smooth monotonous function, which does not resemble the Fano shape. 
Thus, short after the quench only small fluctuations appear near the resonant level and it is not possible to properly fit the curve from formula (\ref{fano}), cf. Figure \ref{fig2b} for $\tilde{t} = 10$ $\hbar/\Gamma_{N}$.  
Therefore, we inspect the evolution of the asymmetry parameter from the moment where local maxima and minima are well distinguished from small disturbances. 
For the set of parameters used in Figure \ref{FanoFitting} ($t^{0}_{12} = 0.4 \Gamma_{N}$), this time was estimated for about $7\ \hbar / \Gamma_{N}$ after the quench.

On four \KK{left} panels of Figure \ref{FanoFitting}, we present the actual shape of conductivity near $eV \simeq \epsilon_2$  (solid red lines) obtained for the same set of parameters as in Figure \ref{AFfit} for times $\tilde{t} = 30,\ 50,\ 80,\ 130$ $\hbar/\Gamma_N$ after quenching (as labelled) and the fitted Fano functions according \KK{to} equation (\ref{fano}) (dashed blue lines). 
The right panel of Figure \ref{FanoFitting} shows the evolution of asymmetry parameter $q$.
The relation between $q$ and the value of $F_{max} \equiv F(\omega_0)$ at the maximum of $F(\omega)$ at $\omega=\omega_0$ is given by
\begin{eqnarray} \label{eq:qvsmax}
     q =\pm\frac{\sqrt{F_{max}-\alpha_{max}}}{\sqrt{\alpha_{max}}},
\end{eqnarray}
where we assumed that $\alpha (\omega)$ is almost not dependent on $\omega$ in the neighbourhood of $\omega_0$ and $\alpha (\omega) \approx  \alpha_{max} \equiv \alpha (\omega_0) $ near $\omega_0$ (cf. also \KK{equation} (10) from Ref. \cite{baranski.zienkiewicz.20}). 
As we showed in Figure \ref{AFfit} with fitting of equation (\ref{function}), the local maxima of the ordinary Fano resonant features evolve exponentially. Consequently, the asymmetry parameter grows exponentially as well, but with different (reduced) characteristic time $\tau_{q}$.
For the same set of parameters as used for the analyses presented in Figure \ref{AFfit}, the corresponding time for the asymmetry parameter was estimated to $\tau_{q}=13.76\ \hbar/\Gamma_N$ (with the fitting function analogous to that given in (\ref{function})).
It differs from the characteristic time for the Fano feature maxima found as $\tau_{F}=17.06\ \hbar/\Gamma_N$, but there is no simple relation between $\tau_{q}$ and $\tau_{F}$ because of the formula (\ref{eq:qvsmax}). 
A small decrease of $q$, for $\tilde{t}> 90\ \hbar/\Gamma_{N}$, is caused by minor oscillations of the local maxima.

Although switching on and off the interdot coupling in considered time scales would be difficult to realize experimentally, effectively the quench protocol presented in this work could be realized electrostatically by gate voltage applied to the side dot (cf., e.g., Refs. \cite{Weymann1,baranski.baranska.21}). 
In such a case, the quench protocol might assume to firstly set the gate voltage, such that energy level of $QD_2$ \KK{lies} beyond considered energy scale. 
At a given moment the gate potential should be changed to a desired energy (within considered energy scale) allowing electron scattering.
Our calculations on this type of \KK{\st{the}} quench protocol indicate that the time scale for formation of both resonances remains the same as establishing an abrupt connection between dots.

\section*{Conclusions}

In this work, we estimated the time required for formation of ordinary Fano resonance and its superconducting response on quantum dot ($QD_{1}$) region upon abrupt connection of additional quantum dot ($QD_2$). 
We found that upon abrupt interdot connection nonequilibrium SC response for scattering has its instantaneous effect pronounced by a high-magnitude asymmetric peak observables in differential conductivity.
Careful inspection of their amplitudes relatively to static results  reveals that stable solution for the anomalous feature develops longer than the feature representing direct scattering. 
The difference in time of this ``saturation'' decreases with increasing of the coupling between dots. 
A comparison of a time scale to typical values of the dot-lead coupling in experiments on quantum dot -- superconductor hybrids shows that time of reaching equilibrium ranges from few to few hundreds of nanoseconds depending on strength of coupling to the Fermi sea and interdot connection. 
Comparing our results to those obtained when $QD_2$ was substituted by one end of topological chain hosting Majorana particles \cite{baranski.baranska.21} we found that establishing of both ordinary and anomalous resonances takes considerably more time than it is required for Majorana mode to leak into the region of $QD$.

\subsection*{Acknowledgments}
We thank Tadeusz Doma\'nski for very fruitful discussions and careful reading of the manuscript.
K.J.K. thanks the Polish National Agency for Academic Exchange for funding in the frame of the Bekker programme (PPN/BEK/2020/1/00184).
K.J.K. is also grateful for the funding from the scholarships of the Minister of Science and Higher Education (Poland) for outstanding young scientists (2019 edition, No. 821/STYP/14/2019).
The equations of motion in the Heisenberg representation have been derived using \textsc{SNEG} library created by R.~\v{Z}itko~\cite{Zitko_2011}.

\subsection*{Author contributions}

J.B. initialized and coordinated the project. 
J.B., T.Z., and M.B. derived the analytic expressions. 
J.B. and K.J.K. performed numerical calculations. 
All authors consulted the obtained results and contributed to the discussions and analysis of the results. 
J.B. and M.B. prepared the first version of the manuscript in interactive discussions with K.J.K.  
All authors reviewed the manuscript. 
J.B. and K.J.K. wrote the paper in its final form. 
All authors accepted it.

\subsection*{Data availability statement}

The data presented in this study are available from the corresponding authors (J.B. -- \url{j.baranski@law.mil.pl} and K.J.K. -- \url{konrad.kapcia@amu.edu.pl}) upon reasonable request. 
All data presented in the paper have been obtained by numerical solving of the set of equations described and/or given in the text of the work.

\subsection*{Competing Interests Statement}

The authors declare no competing interests. The funders had no role in the design of the study; in the collection, analyses, or interpretation of data; in the writing of the manuscript, or in the decision to publish the results.

\end{document}